\documentclass[11pt]{article}
\usepackage{epsfig}
\begin{document}
%
%
%
%
%
\title{{\bf{\Large Statistical Origin of Gravity}}}
\author{
 {\bf {\normalsize Rabin Banerjee}$
$\thanks{E-mail: rabin@bose.res.in}},\, 
 {\bf {\normalsize Bibhas Ranjan Majhi}$
$\thanks{E-mail: bibhas@bose.res.in}}\\
 {\normalsize S.~N.~Bose National Centre for Basic Sciences,}
\\{\normalsize JD Block, Sector III, Salt Lake, Kolkata-700098, India}
\\[0.3cm]
}

\maketitle

\begin{abstract}
    Starting from the definition of entropy used in statistical mechanics we show that it is proportional to the gravity action. For a stationary black hole this entropy is expressed as $S = E/ 2T$, where $T$ is the Hawking temperature and $E$ is shown to be the Komar energy. This relation is also compatible with the generalised Smarr formula for mass.     
\end{abstract}
\vskip 9mm


      There are numerous evidences \cite{Bekenstein:1973ur,Hawking:1974rv,Bardeen:1973gs} which show that gravity and thermodynamics are closely connected to each other. Recently, there has been a growing consensus \cite{Jacobson:1995ab,Kothawala:2007em,Verlinde:2010hp} that gravity need not be interpreted as a fundamental force, rather it is an emergent phenomenon just like thermodynamics and hydrodynamics. The fundamental role of gravity is replaced by thermodynamical interpretations leading to similar or equivalent results. Nevertheless, understanding the entropic or thermodynamic origin of gravity is far from complete since the arguments are more heuristic than concrete and depend upon specific ansatz or assumptions.



    In this paper, using certain basic results derived by us \cite{Banerjee:2008sn,Banerjee:2009pf} in the context of tunneling mechanism, we are able to provide a statistical interpretation of gravity. The starting point is the standard definition of entropy given in statistical mechanics. We show that this entropy gets identified with the action for gravity. Consequently the Einstein equations obtained by a variational principle involving the action can be equivalently obtained by an extremisation of the entropy.

    Furthermore, for a black hole with stationary metric we derive the relation $S=E/2T$, connecting the entropy ($S$) with the Hawking temperature ($T$) and energy ($E$). We prove that this energy corresponds to Komar's expression \cite{Komar:1958wp,Wald:1984rg}. Using this fact we show that the relation $S=E/2T$ is also compatible with the generalised Smarr formula \cite{Smarr:1972kt,Bardeen:1973gs,Gibbons:1976ue}. We mention that this relation was also obtained and discussed in \cite{Padmanabhan:2003pk,Padmanabhan:2009kr}.
 
\vskip 5mm

    We start with the partition function for the space-time with matter field \cite{Gibbons:1976ue},
\begin{eqnarray}
{\cal{Z}} = \int ~ D[g,\Phi] ~  e^{i I[g,\Phi]}
\label{1.01}
\end{eqnarray}
where $I[g,\Phi]$ is the action representing the whole system and $D[g,\Phi]$ is the measure of all field configurations ($g,\Phi$).  
Now consider small fluctuations in the metric ($g$) and the matter field ($\Phi$) in the following form:
\begin{eqnarray}
g=g_0 + {\tilde{g}};\,\,\,\,\ \Phi = \Phi_0 + {\tilde{\Phi}}
\label{1.02}
\end{eqnarray}
where $g_0$ and $\Phi_0$ are the stable background fields satisfying the periodicity conditions and which extremise the action. So they satisfy the classical field equations. Whereas ${\tilde{g}}$ and $\tilde{\Phi}$, the fluctuations around these classical values, are very very small. Expanding $I[g,\Phi]$ around ($g_0,\Phi_0$) we obtain
\begin{eqnarray}
I[g,\Phi] = I[g_0,\Phi_0] + I_2[\tilde{g}] + I_2[\tilde{\Phi}]+{\textrm{higher order terms}}.
\label{1.03}
\end{eqnarray}
The dominant contribution to the path integral (\ref{1.01}) comes from fields that are near the background fields ($g_0,\Phi_0$). Hence one can neglect all the higher order terms. The first term $I[g_0,\Phi_0]$ leads to the usual Einstein equations and gives rise to the standard area law \cite{Gibbons:1976ue}. On the other hand the second and third terms give the contributions of thermal gravitation and matter quanta respectively on the background contribution $I[g_0,\Phi_0]$. They lead to the (logarithmic) corrections to the usual area law \cite{Hawking:1976ja}. Here, since we want to confine ourself within the usual semi-classical regime, we shall neglect these quadratic terms for the subsequent analysis.
Therefore, keeping only the term $I[g_0,\Phi_0]$ in (\ref{1.03}) we obtain the partition function (\ref{1.01}) as \cite{Gibbons:1976ue},
\begin{eqnarray}
{\cal{Z}} \simeq e^{i I[g_0,\Phi_0]}.
\label{1.04}
\end{eqnarray}
Therefore, adopting the standard definition of entropy in statistical mechanics,
\begin{eqnarray}
S = \ln{\cal{Z}} + \frac{E}{T}
\label{stat}
\end{eqnarray}
and using (\ref{1.04}), the entropy of the gravitating system is given by {\footnote{In this paper we have chosen units such that $k_B=G=\hbar=c=1$.}},
\begin{eqnarray}
S= i I[g_0,\Phi_0] + \frac{E}{T}
\label{1.05}
\end{eqnarray}
where $E$ and $T$ are respectively the energy and temperature of the system.

      It may be pointed out that it is possible to interpret (\ref{1.04}) as defining the partition function of an emergent theory without specifying the detailed configuration of the gravitating system. The validity of such an interpretation is borne out by the subsequent analysis.

     In order to get an explicit expression for $E$, let us consider a specific system - a black hole.  Now thermodynamics of a black hole is universally governed by its properties near the event horizon. It is also well understood that near the event horizon the effective theory becomes two dimensional whose metric is given by the two dimensional ($t-r$)- sector of the original metric \cite{Carlip:1998wz,Robinson:2005pd}.
Correspondingly, the left ($L$) and right ($R$) moving (holomorphic) modes are obtained by solving the appropriate field equation using the geometrical (WKB) approximation. Furthermore, the modes inside and outside the horizon are related by the transformations \cite{Banerjee:2008sn,Banerjee:2009pf}: 
\begin{eqnarray}
\phi^{(R)}_{in} &=& e^{-\frac{\pi\omega}{\kappa}} \phi^{(R)}_{out}
\label{mode1}
\\
\phi^{(L)}_{in} &=& \phi^{(L)}_{out}
\label{mode2}
\end{eqnarray}
where ``$\omega$'' is the energy of the particle as measured by an asymptotic observer and ``$\kappa$'' is the surface gravity of the black hole.
In this convention the $L$($R$) moving modes are ingoing (outgoing). Concentrating on the modes inside the horizon, the $L$ mode gets trapped while the $R$ mode tunnels through the horizon and is eventually observed at asymptotic infinity as Hawking radiation \cite{Banerjee:2008sn,Banerjee:2009pf}.
The probability of this ``$R$'' mode, to go outside, as measured by the outside observer is given by 
\begin{eqnarray}
P^{(R)}=\Big|\phi^{(R)}_{in}\Big|^2 = \Big|e^{-\frac{\pi\omega}{\kappa}} \phi^{(R)}_{out}\Big|^2 = e^{-\frac{2\pi\omega}{\kappa}}
\label{mode3}
\end{eqnarray} 
where, in the second equality, (\ref{mode1}) has been used. This is essential since the measurement is done from outside and hence $\phi^{(R)}_{in}$ has to be expressed in favour of $\phi^{(R)}_{out}$. Therefore the average value of the energy, measured from outside, is given by,
\begin{eqnarray}
<\omega> = \frac{\int_0^{\infty}~d\omega~\omega ~e^{-\frac{2\pi\omega}{\kappa}}}{\int_0^{\infty}~d\omega~ e^{-\frac{2\pi\omega}{\kappa}}} = T
\label{mode4}
\end{eqnarray}
where $T=\kappa/2\pi$ is the temperature of the black hole \cite{Banerjee:2009pf}. Therefore if we consider that the energy $E$ of the system is encoded near the horizon and the total number of pairs created is $n$ among which this energy is distributed, then we must have,
\begin{eqnarray}
E=nT
\label{new1}
\end{eqnarray}
where only the $R$ mode of the pair is significant.

     Now to proceed further, it must be realised that the effective two dimensional curved metric can always be embedded in a flat space which has exactly two space-like coordinates. This is a consequence of a modification in the original GEMS (globally embedding in Minkowskian space) approach of \cite{Deser:1998xb} and has been elaborated by us in \cite{Banerjee:2010ma}. Hence we may associate each $R$ mode with two degrees of freedom. Then the total number of degrees of freedom for $n$ number of $R$ modes is $N=2n$. Hence, 
from (\ref{new1}), we obtain the energy of the system as
\begin{eqnarray}
E = \frac{1}{2}NT.
\label{1.07}
\end{eqnarray}
As a side remark, it may be noted that (\ref{1.07}) can be interpreted as a consequence of the usual law of equipartition of energy. For instance, if we consider that the energy $E$ is distributed equally over each degree of freedom, then 
(\ref{1.07}) implies that each degree of freedom should contain an energy equal to $T/2$, which is nothing but the {\it equipartition law of energy}. The fact that the energy is equally distributed among the degrees of freedom may be understood from the symmetry of two space-like coordinates ($z^1\longleftrightarrow z^2$) such that the metric is unchanged \cite{Banerjee:2010ma}. In our subsequent analysis, however, we only require (\ref{1.07}) rather than its interpretation as the law of equipartition of energy.

       Now since there are $N$ number of degrees of freedom in which all the information is encoded, the entropy ($S$) of the system must be proportional to $N$. Hence using (\ref{1.05}) we obtain
\begin{eqnarray}
N = N_0 S = N_0 (i I[g_0,\Phi_0]  + \frac{E}{T}),
\label{1.06}
\end{eqnarray} 
where $N_0$ is a proportionality constant, which will be determined later.
Substituting the value of $N$ from (\ref{1.07}) in (\ref{1.06}) we obtain the expression for the energy of the system as
\begin{eqnarray}
E=\frac{N_0}{2 - N_0}iT I[g_0,\Phi_0].
\label{1.08}
\end{eqnarray}
This shows that in the absence of any fluctuations, the energy of a system is actually given by the classical action representing the system. In the following we shall use this expression to find the energy of a stationary black hole. Before that let us substitute the value of $I[g_0,\Phi_0]$ from (\ref{1.08}) in (\ref{1.05}). This immediately leads to a simple relation between the entropy, temperature and energy of the black hole:
\begin{eqnarray}
S=\frac{2E}{N_0T}.
\label{entropy1}
\end{eqnarray}
Now in order to fix the value of ``$N_0$'' we consider the simplest example, the Schwarzschild black hole for which the entropy, energy and temperature are given by,
\begin{eqnarray}
S = \frac{A}{4} = 4\pi M^2, \,\,\  E = M,  \,\,\ T = \frac{1}{8\pi M},
\label{Sch1}
\end{eqnarray}
where ``$M$'' is the mass of the black hole.
Substitution of these in (\ref{entropy1}) leads to $N_0 = 4$.

     At this point we want to make a comment on the value of $N_0$. According to standard statistical mechanics one would have thought that $1/N_0 = \ln c$, where $c$ is an integer. Whereas to keep our analysis consistent with semi-classical area law, we obtained $c=e^{1/4}$, which is clearly not an integer. Indeed, any departure from this value of $N_0$ would invalidate the semi-classical area law and hence our analysis. Such a disparity is not peculiar to our approach and has also occurred elsewhere \cite{Paddy}. This may be due to the fact that our analysis is confined within the semi-classical regime, which is valid for large degrees of freedom. In this regime, it is not obvious that a semi-classical computation can reproduce $c$ to be an integer. Furthermore, the above value of $N_0$ is still valid even for very small number of degrees of freedom ($N$), where this semi-classical calculation is unjustified. This also happens in the semi-classical computation of the entropy spectrum of a black hole \cite{Maggiore:2007nq}. The entropy spectrum is found there to be $S=2\pi N$ rather than $S=N\ln c$ and this discrepancy is identified with the semi-classical approximation. A possible way to resolve such disagreement from standard statistical mechanics may be the full quantum theoretical computation of the number of microstates which is beyond the scope of the present paper.

     Finally, putting back  $N_0=4$ in (\ref{entropy1}) we obtain,
\begin{eqnarray}
S=\frac{E}{2T}.
\label{entropy}
\end{eqnarray}
Before discussing the significance and implications of this relation, we observe that substituting the value of $E$ from (\ref{entropy}) in (\ref{1.08}) with $N_0 = 4$, we obtain
\begin{eqnarray}
S = -iI[g_0,\Phi_0].
\label{einstein}
\end{eqnarray}
Consequently, extremisation of entropy leads to Einstein's equations.

    The relation (\ref{entropy}) is significant for various reasons which will become progressively clear. It is valid for all black hole solutions in Einstein gravity with appropriate identifications consistent with the area law. Here $S$ and $T$ are easy to identify. These are, respectively, the entropy and Hawking temperature of the black hole. Since energy is one of the most diversely defined entities in general theory of relativity, special care is needed to identify $E$ in (\ref{entropy}). We now show that this $E$ corresponds to Komar's definition \cite{Komar:1958wp,Wald:1984rg}. Simplifying (\ref{1.08}) using $N_0 = 4$ and $T=\kappa/2\pi$, we obtain,
%
%
%
\begin{eqnarray}
E = -\frac{i\kappa I[g_0,\Phi_0]}{\pi}.
\label{1.09}
\end{eqnarray}
The classical action $I[g_0,\Phi_0]$ has already been calculated in \cite{Gibbons:1976ue}. The result is,
\begin{eqnarray}
I[g_0,\Phi_0] = 2i\pi\kappa^{-1}\Big[\frac{1}{16\pi}\int_{\Sigma}R\xi^ad\Sigma_a + \int_{\Sigma}(T_{ab} - \frac{1}{2}Tg_{ab})\xi^bd\Sigma^a - \frac{1}{16\pi}\int_{\cal{H}}\epsilon_{abcd}\nabla^c\xi^d\Big],
\label{1.10}
\end{eqnarray}
where $\xi^a\partial/\partial x^a = \partial/\partial t$ is the time translation Killing vector and $\Sigma$ is the space-like hypersurface whose boundary is given by ${\cal{H}}$. Here $T_{ab}$ is the energy-momentum tensor of the matter field whose trace is given by $T$.  
Now for a stationary geometry, $\xi^a\nabla_aR=0$ \cite{Carroll:2004st}. Hence for a volume ${\cal{A}}$, we have
\begin{eqnarray}
0 = \int_{\cal{A}}\xi^a\nabla_aR d{\cal{A}} = \int_{\cal{A}}\Big[\nabla_a(\xi^a R) -(\nabla_a\xi^a)R\Big] d{\cal{A}} = \int_{\cal{A}}\nabla_a(\xi^a R)d{\cal{A}}
\label{1.11}
\end{eqnarray}
where in the last step the Killing equation $\nabla_a\xi_b+\nabla_b\xi_a=0$ has been used.
Finally, the Gauss theorem yields,
\begin{eqnarray}
\int_{\Sigma}\xi^a Rd{\Sigma_a} = 0.
\label{1.12}
\end{eqnarray}
Using this in (\ref{1.10}) we obtain,
\begin{eqnarray}
I[g_0,\Phi_0] = 2i\pi\kappa^{-1}\Big[\int_{\Sigma}(T_{ab} - \frac{1}{2}Tg_{ab})\xi^bd\Sigma^a - \frac{1}{16\pi}\int_{\cal{H}}\epsilon_{abcd}\nabla^c\xi^d\Big].
\label{1.13}
\end{eqnarray}
Substituting this in (\ref{1.09}) we obtain the expression for the energy of the gravitating system as
\begin{eqnarray}
E = 2\int_{\Sigma}(T_{ab} - \frac{1}{2}Tg_{ab})\xi^bd\Sigma^a - \frac{1}{8\pi}\int_{\cal{H}}\epsilon_{abcd}\nabla^c\xi^d
\label{1.14}
\end{eqnarray}
which is the Komar expression for energy \cite{Komar:1958wp,Wald:1984rg} corresponding to the time translation Killing vector. Similarly, if there is a rotational Killing vector, then there must be a Komar expression for rotational energy \cite{Carroll:2004st,Katz} and the total energy will be their sum.

    Incidentally, (\ref{entropy}) was obtained earlier in \cite{Padmanabhan:2003pk} for static space-time and its implications were discussed in \cite{Padmanabhan:2009kr}. However a specific `ansatz' for entropy compatible with the area law was taken and, more importantly, the Komar energy expression was explicitly used as an input in the derivation. Hence our analysis is completely different, since we do not invoke any ansatz for the entropy; neither is the Komar expression required at any stage. Rather we prove its occurence in the relation (\ref{entropy}).

     As an explicit check of (\ref{entropy}) for different black hole solutions, we consider a couple of examples. Take the Reissner-Nordstr$\ddot{\textrm{o}}$m (RN) black hole. In this case the entropy and temperature are given by,
\begin{eqnarray}
S = \pi r_+^2, \,\,\,\ T = \frac{r_+ - r_-}{4\pi r_+^2}; \,\,\ r_{\pm} = M \pm \sqrt{M^2 - Q^2}
\label{RN1}
\end{eqnarray}
where ``$Q$'' is the charge of the black hole.
Substitution of these in (\ref{entropy}) yields,
\begin{eqnarray}
E = M - \frac{Q^2}{r_+},
\label{RN2}
\end{eqnarray}
which is the Komar energy of RN black hole \cite{Banerjee:2009tz}.

     Next we consider the Kerr black hole for which the entropy and temperature are respectively,
\begin{eqnarray}
S &=& \pi (r_+^2 + a^2), \,\,\,\ T = \frac{r_+ - r_-}{4\pi (r_+^2 + a^2)};
\nonumber
\\
r_{\pm} &=& M \pm \sqrt{M^2 - a^2}, \,\,\,\ a = \frac{J}{M}.
\label{Kerr1}
\end{eqnarray}
Here ``$J$'' is the angular momentum of the black hole. Substituting (\ref{Kerr1}) in (\ref{entropy}) we obtain,
\begin{eqnarray}
E = M - 2J\Omega
\label{Kerr2}
\end{eqnarray}
which is the total Komar energy for Kerr black hole \cite{Dadich,Banerjee:2009tz}. Here $\Omega = \frac{a}{r_+^2 + a^2}$ is the angular velocity at the event horizon $r=r_+$.

   We thus find that, in all cases where $S$, $E$, $T$ are known, they satisfy (\ref{entropy}) apart from the area law. In fact, it is possible to take (\ref{entropy}) as the defining relation for the Komar energy in those examples where its direct calculation from (\ref{1.14}) is difficult. Such an instance is provided by the Kerr-Newman black hole. Its Komar energy, as far as we aware, is not known in closed form. However the entropy and temperature of Kerr-Newman black hole are given by,
\begin{eqnarray}
S= \pi (r_+^2 + a^2); \,\,\,\ T=\frac{r_+ - r_-}{4\pi(r_+^2 + a^2)}
\label{KN1}
\end{eqnarray}
where
\begin{eqnarray}
r_{\pm} = M \pm \sqrt{M^2-Q^2-a^2}; \,\,\,\ a = \frac{J}{M}.
\label{KN2}
\end{eqnarray}
Now substituting (\ref{KN1}) in (\ref{entropy}) and then using (\ref{KN2}) we obtain the total Komar energy of Kerr-Newman black hole:
\begin{eqnarray}
E = \sqrt{M^2 - Q^2 - a^2} = M - \frac{Q^2}{r_+} - 2J\Omega\Big(1-\frac{Q^2}{2Mr_+}\Big) = M - QV - 2J\Omega,
\label{KN3}
\end{eqnarray}  
where $\Omega = \frac{a}{r_+^2 + a^2}$ is the angular velocity at the event horizon and $V = \frac{Q}{r_+} - \frac{QJ\Omega}{M r_+}$.
This value exactly matches with the direct evaluations of Komar expressions for energies within the first order approximation \cite{Katz,Dadich,Banerjee:2009tz}. It is also reassuring to note that the definition of $M$ following from (\ref{entropy}) and (\ref{KN3}) reproduces the generalised Smarr formula \cite{Smarr:1972kt,Bardeen:1973gs,Gibbons:1976ue},
\begin{eqnarray}
\frac{M}{2} = \frac{\kappa A}{8\pi} + \frac{VQ}{2} + \Omega J.
\label{Smarr}
\end{eqnarray}
\vskip 5mm


     In this paper we have further clarified the possibility of considering gravity as an emergent phenomenon. Taking the standard definition of entropy from statistical mechanics we were able to show the equivalence of entropy with the action. Consequently, extremisation of the action leading to Einstein's equations is equivalent to the extremisation of the entropy. We derived the relation $S = E/2T$ for stationary black holes with $S$ and $T$ being the entropy and Hawking temperature. The nature of energy $E$ appearing in this relation was clarified. It was proved to be Komar's expression valid for stationary asymptotically flat space-time. An explicit check of $S=E/2T$ was done for all black hole solutions of Einstein gravity. This relation was also seen to reproduce the generalised mass formula of Smarr \cite{Smarr:1972kt,Bardeen:1973gs,Gibbons:1976ue}. In this sense the Smarr formula can be interpreted as a thermodynamic relation further illuminating the emergent nature of gravity. As a final remark we feel that although our results were derived for Einstein gravity, the methods are general enough to include other possibilities like higher order theories. 

\vskip 9mm

{\bf{Acknowledgement:}}\linebreak
The authors thank Mr. S. K. Modak for useful discussions.


\begin{thebibliography}{99}
\bibitem{Bekenstein:1973ur}
  J.~D.~Bekenstein,
  Phys.\ Rev.\  D {\bf 7}, 2333 (1973).
\bibitem{Hawking:1974rv}
  S.~W.~Hawking,
  Nature {\bf 248}, 30 (1974).\\
  S.~W.~Hawking,
  Commun.\ Math.\ Phys.\  {\bf 43}, 199 (1975)
  [Erratum-ibid.\  {\bf 46}, 206 (1976)].
\bibitem{Bardeen:1973gs}
  J.~M.~Bardeen, B.~Carter and S.~W.~Hawking,
  Commun.\ Math.\ Phys.\  {\bf 31}, 161 (1973).
\bibitem{Jacobson:1995ab}
  T.~Jacobson,
  Phys.\ Rev.\ Lett.\  {\bf 75}, 1260 (1995)
  [arXiv:gr-qc/9504004].
\bibitem{Kothawala:2007em}
  T.~Padmanabhan,
  arXiv:0911.5004 [gr-qc] and references therein.
\bibitem{Verlinde:2010hp}
  E.~P.~Verlinde,
  arXiv:1001.0785 [hep-th].
\bibitem{Banerjee:2008sn}
  R.~Banerjee and B.~R.~Majhi,
  Phys.\ Rev.\  D {\bf 79}, 064024 (2009)
  [arXiv:0812.0497 [hep-th]].\\
  R.~Banerjee and B.~R.~Majhi,
  Phys.\ Lett.\  B {\bf 675}, 243 (2009)
  [arXiv:0903.0250 [hep-th]].
\bibitem{Banerjee:2009pf}
  R.~Banerjee, B.~R.~Majhi and E.~C.~Vagenas,
  Phys.\ Lett.\  B {\bf 686}, 279 (2010)
  [arXiv:0907.4271 [hep-th]].
\bibitem{Komar:1958wp}
  A.~Komar,
  Phys.\ Rev.\  {\bf 113}, 934 (1959).
\bibitem{Wald:1984rg}
  R.~M.~Wald,
  ``General Relativity,''
{\it  Chicago, Usa: Univ. Pr. ( 1984) 491p}. 
\bibitem{Smarr:1972kt}
  L.~Smarr,
  Phys.\ Rev.\ Lett.\  {\bf 30}, 71 (1973)
  [Erratum-ibid.\  {\bf 30}, 521 (1973)].
\bibitem{Gibbons:1976ue}
  G.~W.~Gibbons and S.~W.~Hawking,
  Phys.\ Rev.\  D {\bf 15}, 2752 (1977).
\bibitem{Padmanabhan:2003pk}
  T.~Padmanabhan,
  Class.\ Quant.\ Grav.\  {\bf 21}, 4485 (2004)
  [arXiv:gr-qc/0308070].
\bibitem{Padmanabhan:2009kr}
  T.~Padmanabhan,
  arXiv:0912.3165 [gr-qc].
\bibitem{Hawking:1976ja}
  S.~W.~Hawking,
  Commun.\ Math.\ Phys.\  {\bf 55}, 133 (1977).
\bibitem{Carlip:1998wz}
  S.~Carlip,
  Phys.\ Rev.\ Lett.\  {\bf 82}, 2828 (1999)
  [arXiv:hep-th/9812013].
\bibitem{Robinson:2005pd}
  S.~P.~Robinson and F.~Wilczek,
  Phys.\ Rev.\ Lett.\  {\bf 95}, 011303 (2005)
  [arXiv:gr-qc/0502074].
\bibitem{Deser:1998xb}
  S.~Deser and O.~Levin,
  Phys.\ Rev.\  D {\bf 59}, 064004 (1999)
  [arXiv:hep-th/9809159].
\bibitem{Banerjee:2010ma}
  R.~Banerjee and B.~R.~Majhi,
  arXiv:1002.0985 [gr-qc] (Phys. Lett. B, In press).
\bibitem{Carroll:2004st}
  S.~M.~Carroll,
  ``Spacetime and geometry: An introduction to general relativity,''
{\it  San Francisco, USA: Addison-Wesley (2004) 513 p}.
\bibitem{Katz}
  J.~Katz,
  Class.\ Quant.\ Grav. {\bf 2}, 423 (1985).
\bibitem{Paddy} See, for instance, the discussion in section 2 of \cite{Padmanabhan:2009kr}.
\bibitem{Maggiore:2007nq}
  M.~Maggiore,
  Phys.\ Rev.\ Lett.\  {\bf 100}, 141301 (2008)
  [arXiv:0711.3145 [gr-qc]]. See, in particular, the last section.
\bibitem{Dadich}
  R.~Kulkarni, V.~Chellathurait and N.~Dadhich,
  Class.\ Quant.\ Grav. {\bf 5}, 1443 (1988)\\
  V.~Chellathurait and N.~Dadhich,
  Class.\ Quant.\ Grav. {\bf 7}, 361 (1990).\\
  N.~Dadhich,
  Phys.\ Lett. A {\bf 98}, 103 (1983).
\bibitem{Banerjee:2009tz}
  R.~Banerjee and S.~K.~Modak,
  JHEP {\bf 0905}, 063 (2009)
  [arXiv:0903.3321 [hep-th]].
\end{thebibliography}
\end{document}